\documentclass[a4paper]{article}
\usepackage{graphicx,epsfig,rotating}
\begin{document}
 
\newcommand{\ra}{$\;\rightarrow$}
\def\la{\mathrel{\mathpalette\fun <}}
\def\ga{\mathrel{\mathpalette\fun >}}
\def\fun#1#2{\lower3.6pt\vbox{\baselineskip0pt\lineskip.9pt
\ialign{$\mathsurround=0pt#1\hfil##\hfil$\crcr#2\crcr\sim\crcr}}} 

\begin{titlepage}

\vspace*{15mm}
\begin{flushright}{\large\bf May 17, 1999}\end{flushright}

\vspace*{20mm}
 
\begin{center}

{\LARGE\bf Observability of $B^{0}_{(d)s} \rightarrow \mu \mu$ decay with
the CMS detector}

\vspace*{15mm}

  \begin{center}
   A.~Nikitenko~$^{a)}$ \\
   HIP, Helsinki, Finland \\
   A.~Starodumov~$^{b)}$ \\
   Universit$\grave{a}$ di Pisa e Sezione dell' INFN, Pisa, Italy \\
   N.~Stepanov~$^{c)}$ \\ 
   CERN, Geneva, Switzerland

\end{center}
\end{center}

\vspace{20mm}

\begin{abstract}
We have updated our previous study on the possibility to observe the rare 
$B^{0}_{S} \rightarrow \mu \mu$ decay with detailed simulation of the
TDR version of the CMS detector. The full simulation and reconstruction 
procedure has been done both for the signal and background. We improved our 
previous results on expected sensitivity using more sophisticated algorithms 
for the track and vertex reconstruction and tighter selection criteria.
\end{abstract} 

\vspace{50mm}
\hspace*{-8mm} \hrulefill \hspace{60mm} \hfill \\
$^{a)}$  Email: Alexandre.Nikitenko@cern.ch\\
$^{b)}$ Email: Andrei.Starodumov@cern.ch \\  
$^{c)}$ Email: Nikita.Stepanov@cern.ch \\

\end{titlepage}


\section{Introduction}

The purpose of this study is to update our earlier study \cite{bs_old} on the 
possibility to observe the rare $B^{0}_{S} \rightarrow \mu \mu$ decay with 
the CMS detector. As well known, this channel with a very small Standard
Model branching ratio is a sensitive probe of a ``new physics'' affecting
FCNC \cite{br_rat}, and only the large b production rate of a hadron collider
possibly allows to obtain the needed level of sensitivity. The reasons which 
might modify the previous results are the following (for details see the 
next chapters) :

- detailed simulation of the TDR version of the CMS detector, tracker in
particular;

- more sophisticated algorithms for track and vertex reconstruction and 
tighter selection criteria;

- new version of the PYTHIA Monte-Carlo generator \cite{pythia};

- improved statistical precision of the results due to a much bigger sample 
of generated events;
 
- new theoretical estimate of the branching ratio of the decay in the
Standard Model $(3.5 \pm 1.0) \times 10^{-9}$ \cite{br_rat};

In the previous study \cite{bs_old} we established that the dominant 
background is due to direct muons from $b \bar{b}$ pairs produced 
through the gluon splitting mechanism. All other sources of background are 
found  to be about
one order of magnitude smaller than this one. In the present work we thus 
concentrated only on this source of background. The higher level trigger and
kinematics selection criteria were taken as in \cite{bs_old}:

\begin{equation}
p_{t}^{\mu} > 4.3~GeV,~~~|\eta| < 2.4
\end{equation}
\begin{equation}
0.4 < \Delta R_{\mu \mu} < 1.2,~~~p_{t}^{\mu \mu} > 12~GeV
\end{equation}

where $R_{\mu \mu}$ is the distance between the two muons in $\eta$, 
$\phi$ space and $p_{t}^{\mu \mu}$ is the transverse momentum of the di-muon 
pair. We assume that there is no significant loss of dimuon events at the
first trigger level. This is justified for running at luminosity 
$10^{33}cm^{-2}s^{-1}$ but should be examined in more detail when 
approaching $10^{34}cm^{-2}s^{-1}$. 
To suppress the background we have exploited as in our previous study 
the good dimuon mass resolution of CMS, the precise two-muon vertex 
reconstruction, and isolation criteria with the tracker and calorimeters.

We have also investigated sensitivity to  
$B^{0}_{d} \rightarrow \mu \mu$ decays.

\section{Event generation}

\subsection{Kinematics and cross-sections}

Background events have been generated by PYTHIA5.7 with default CTEQ2L 
structure functions and the default choice of the fragmentation function,
which is
the Lund symmetric fragmentation function modified for heavy endpoint
quarks (see references in \cite{pythia}). Pairs of $b\bar{b}$ quarks have
been extracted from MSEL=1 QCD 2 $\rightarrow$ 2 processes where gluons
are produced: process 28 (f+g $\rightarrow$ f+g) and 68 
(g+g $\rightarrow$ g+g). Data selection kinematics cuts (1), (2) on muons 
have been applied at the generation level. We have generated 10000 of di-muon 
background events passing through these cuts. 

The normalisation has been done on the $b\bar{b}$ production cross-section 
obtained by counting the number of events where at least one $b\bar{b}$ 
pair was produced. We give in Table 1 the PYTHIA output on the 
cross-sections of MSEL=1 subprocesses and cross-section of $b\bar{b}$ pairs 
production with the processes 11-13, 53 and 28,68. The total fraction of  
events with $b\bar{b}$ pairs is $7.4 \times 10^{-3}$ and the corresponding 
cross-section is 409 $\mu$b. According to the data in Table 1, we have 
normalised the background from gluon splitting on 282 $\mu$b. In the previous
study we normalised gluon splitting background on 370 $\mu$b given by
PYTHIA5.6 with EHLQ1 structure functions (PYTHIA5.6 default).
   
Signal $B^{0}_{S} \rightarrow \mu \mu$ events have been extracted from all
MSEL=1 subprocesses and are present therefore a mixture of gluon 
splitting and gluon fusion production mechanisms.

We estimate that after trigger and kinematic selections the number
of signal and background events, for one year running at luminosity
$L=10^{33}cm^{-2}s^{-1}$ (i.e. $10^{4}pb^{-1}$), is 66 and 
$2.9 \times 10^{7}$ respectively.

We checked that choosing the Peterson fragmentation function instead of the
default one does not make a difference in the spectra of B-hadrons and charged 
particles around the B-hadron and therefore does not affect the efficiency of 
the kinematics selections and isolation criteria (see below). 
In Figure~\ref{fragm_func} are shown the spectra of the B-hadron 
(Figure~\ref{fragm_func}a) and of charged 
particles (Figure~\ref{fragm_func}b) with $p_{t}>0.9$ GeV (not including 
decay products of B-hadron) in cone $\Delta R <$0.4 around the B-hadron for 
the default choice of the fragmentation function (solid line) and the 
Peterson fragmentation 
function \cite{peterson} (dashed line). For this plot, $b \bar{b}$ pairs have 
been generated by a gluon fusion mechanism $gg \rightarrow b\bar{b}$ with 
$\hat{p_{t}} >$4 GeV. There is almost no difference between the two choices 
of the fragmentation function.
 
\begin{verbatim}

Table 1. Cross-sections for MSEL=1 subprocesses and corresponding
         cross-sections for bb pair production in these processes.
  -------------------------------------------------------------------------
         Subprocesses             Cross-section     Cross-section of 
                                     (mb)         bb pairs production (mb)
  -------------------------------------------------------------------------
   I  11 f + f' -> f + f' (QCD) I  1.528E+00   I
   I  12 f + f~ -> f' + f~'     I  2.209E-02   I        0.002
   I  13 f + f~ -> g + g        I  2.086E-02   I        
  -------------------------------------------------------------------------
   I  53 g + g -> f + f~        I  8.553E-01   I        0.125
  -------------------------------------------------------------------------
   I  28 f + g -> f + g         I  1.659E+01   I
   I  68 g + g -> g + g         I  3.621E+01   I        0.282
  -------------------------------------------------------------------------
   I  All included subprocesses I  5.522E+01   I        0.409       
  -------------------------------------------------------------------------

\end{verbatim}

\subsection{CMS detector simulation}

Detector simulation has been done with the CMSIM package \cite{cms114} which
is based on GEANT3 and simulates properly the response of the CMS subdetectors
used for the reconstruction. The TDR designs for the calorimetry and 
tracker are taken. Two options of the tracker geometry have been simulated, 
corresponding to the low and high luminosity running. The important
difference between these two designs is the position of the pixel vertex 
detector. The radii of the two barrel pixel cylinders are 4 and 7 cm 
for the low luminosity design, and 7 and 11 cm for the high luminosity case.
The low luminosity configuration provides a better vertex reconstruction 
precision over the entire tracker acceptance \cite{tracker_tdr}.   
    
Both signal and background events were
passed through the detailed detector simulation and reconstruction procedure 
to obtain
the efficiency of the selection criteria. Energy thresholds on particle 
tracking is taken as 100 KeV (1 MeV) for electrons and photons, and 
1 MeV (10 MeV) for hadrons in the Tracker (Calorimetry).  

\section{Event selections}

\subsection{Track and vertex reconstruction}

The CM\_FKF track finder \cite{tracker_tdr} was used to reconstruct
all tracks with $p_{t} \geq 0.9$ GeV, $\mid \eta \mid \leq 2.5$ with at 
least 6 hits and $\chi^2 / ndf \leq 7$.
In these conditions, the global 
track reconstruction efficiency is about 91 \% both for signal and
background samples. After track reconstruction the GVF vertex finder
\cite{gvf} was implemented. The GVF allows in principle to reconstruct all 
"interesting"
vertices in a given event, including primary, secondary and two prong
particle decay vertices. However, in our particular case, a simplified
version of the algorithm is used, tuned to reconstruct just the dimuon 
secondary vertices. A given reconstructed track was considered to 
be a muon and was taken into account in the vertex reconstruction,
if it had at least two pixel hits and was associated with the true
Monte Carlo muon, i.e. no realistic tracker-muon system matching has been 
implemented yet. We consider that the efficiency for this matching should be
very high as the b-jets are relatively soft i.e. open and of moderate
multiplicity while the muons are typically will be of $p_{t}^{\mu}\geq$ 6 GeV. 
To understand the vertex quality cuts 
used below one has to keep in
mind that the GVF is a two-step procedure. In this case it creates first
the vertex seeds
trying all dimuon combinations. A given track pair is accepted as a secondary
vertex seed if the minimal 3D distance (called M2D below) 
between the two spirals is small enough. At this stage some tracks can 
belong, in principle, to several vertex seeds. At a second stage of the 
algorithm all accepted seeds are processed by the dedicated algorithm, 
which tries to solve ambiguities and to fit simultaneously the 
vertex positions and the track parameters.

Several event samples have been used to optimise the selection criteria
and the signal to background ratio. 
     
\begin{enumerate}

\item $\sim$ 400 signal events.

\item 3000 signal events with muons only. Other particles in the event 
were not propagated through the detector. This sample has been used to tune 
the vertex reconstruction and selection criteria.

\item 10000 background events.

\item 10000 same background events, but bearing only the muons.

\end{enumerate}

The samples with muons only have been used mostly to evaluate the high 
luminosity scenario, since the processing of a full event (bunch crossing),
with pile-up, takes enormous CPU and memory resources at the moment.

\begin{figure}[htbp]
\centering
\resizebox{180mm}{180mm}
{\includegraphics{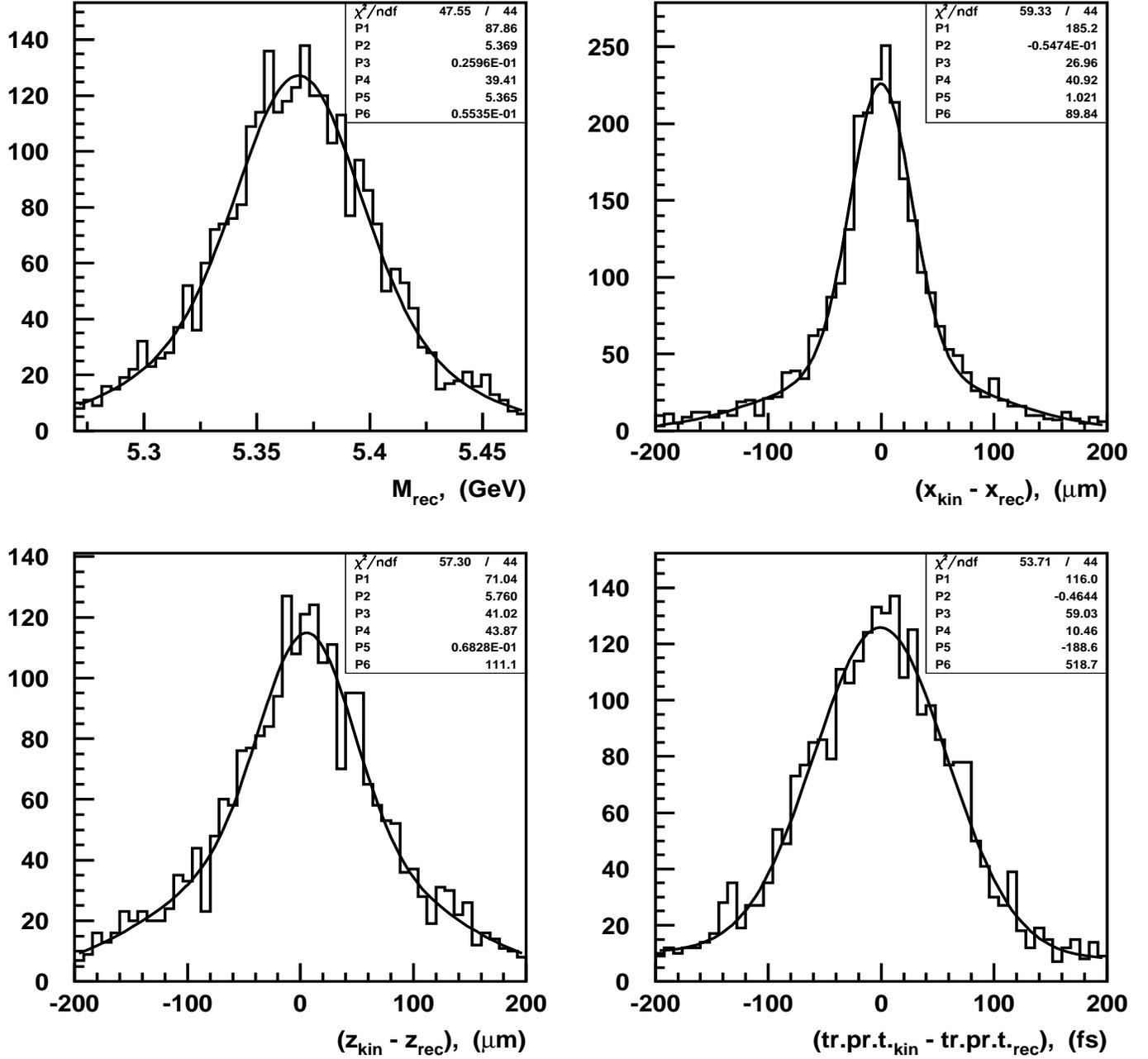}}
\caption
{Signal events. Dimuon mass resolution, resolution in X/Y, resolution in Z,
time resolution}
\label{global_var}
\end{figure}

The basic variables which characterise the performance of the low luminosity 
tracker configuration are shown in Figure~\ref{global_var}. In this figure
one can see dimuon mass resolution, secondary vertex resolution in X/Y and 
Z coordinates and proper time resolution.

\subsection{Selections criteria based on the tracker}

\textbf{i) mass cut}

The first background suppression criterion is a 
$B^{0}_{s} \rightarrow \mu \mu$ mass cut against the continuum dimuon
background population. To search for signal events we have chosen a di-muon 
mass window of $\pm$ 40 MeV centred on the $B_{S}$ mass of 5.369 GeV. 
In Figure~\ref{isol_mass_corr}a 
the di-muon mass spectrum for the background is shown after the trigger and 
initial kinematics selections. The 80 MeV mass window is also shown in 
the same figure. Only $1.1 \pm 0.1~\%$ of background events is retained
in this mass window. As full simulations have shown, the high luminosity
tracker configuration provides nearly the same di-muon mass resolution.
For high luminosity running we thus expect the same selectivity of 
the dimuon mass cut both for the signal and background as for the 
low luminosity case.

\textbf{ii) isolation}

$B^{0}_{s} \rightarrow \mu \mu$ decays produced in soft b-jets are
semi-isolated and significantly more isolated than dimuons from accidental
$g \rightarrow b(\rightarrow \mu) \bar{b}(\rightarrow \mu)$ associations.
In our previous study \cite{bs_old} it was found that the rejection 
factor based on the tracker isolation depends strongly on the lower cutoff 
on the transverse momentum of the charged particles in the isolation cone 
around the di-muon momentum. The detailed tracker pattern recognition 
studies show that charged tracks with transverse momenta above 0.9 GeV can 
be reconstructed with an efficiency exceeding 90 \% within the tracker 
acceptance \cite{tracker_tdr}.
In this work we have used a slightly different tracker
isolation definition than in \cite{bs_old}. We required no charged
tracks with $p_{t} >$ 0.9 GeV in a cone 
$\Delta R~=~0.5 \times \Delta R_{\mu \mu}~+~0.4$ around the di-muon 
momentum direction. Such a criterion gives an efficiency of 0.49 for
the signal and $(3.0 \pm 0.2) \times 10^{-2}$ for the background in 
conditions of low luminosity running.

\begin{figure}[htbp]
\centering
\resizebox{180mm}{180mm}
{\includegraphics{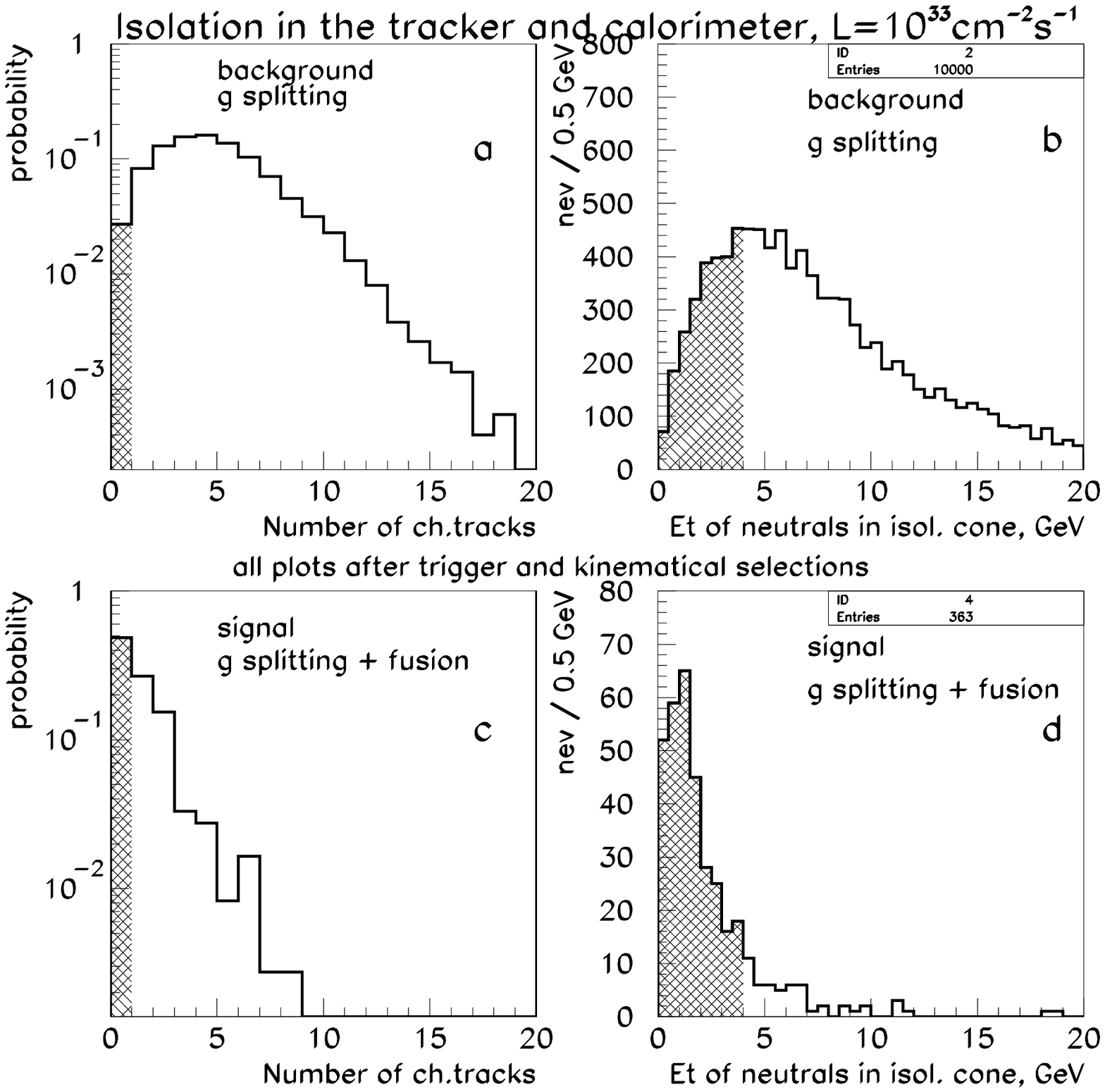}}
\caption
{Tracker isolation parameter (see text) for the background (a) and signal (b),
calorimeter isolation energy for the background (b) and signal (d) calculated
with particle level simulation (as explained in the text) for the case of
$L=10^{33}cm^{-2}s^{-1}$.}
\label{isol_ll}
\end{figure}

\begin{figure}[htbp]
\centering
\resizebox{160mm}{160mm}
{\includegraphics{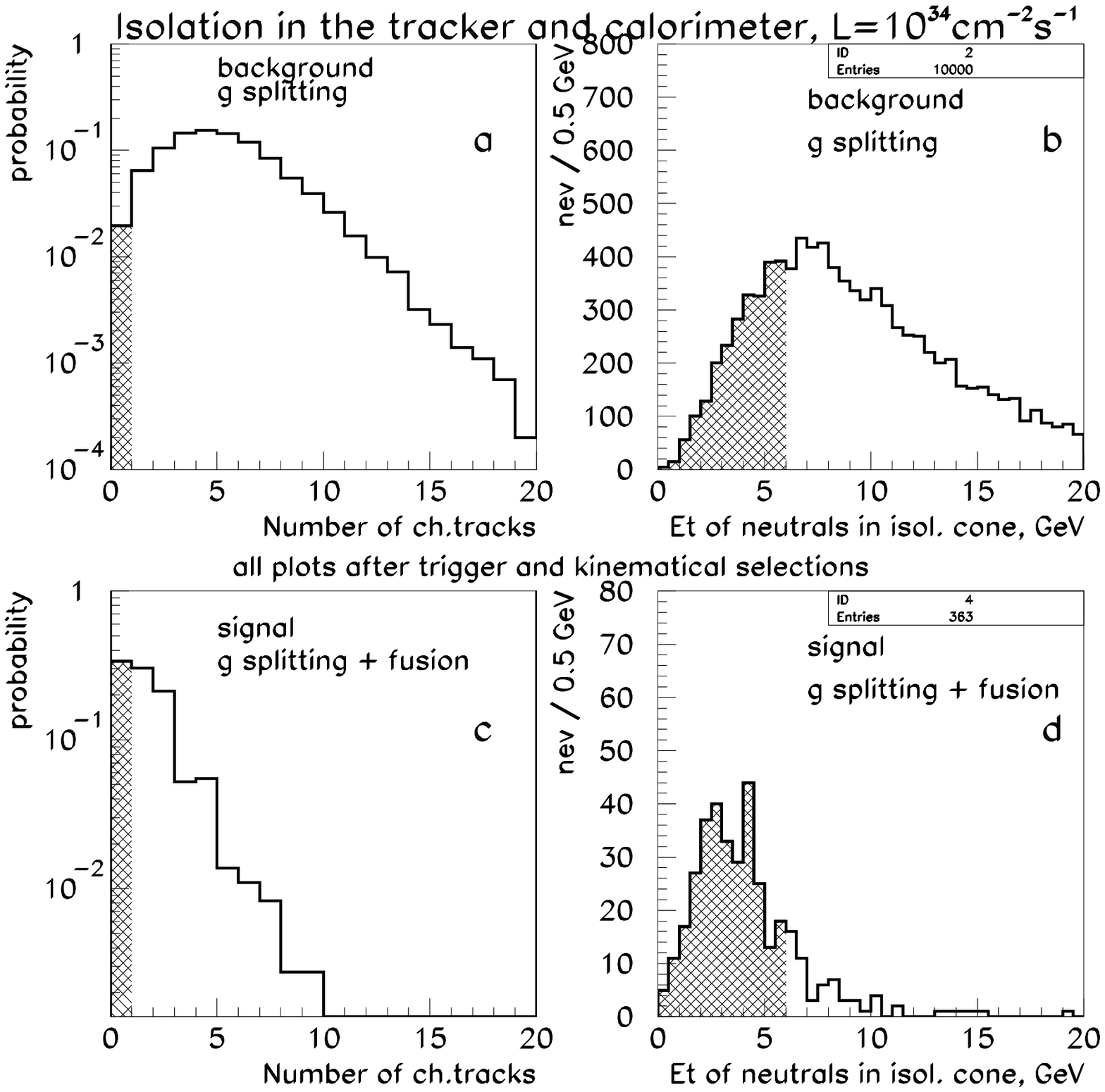}}
\caption
{Tracker isolation parameter (see text) for the background (a) and signal (b),
calorimeter isolation energy for the background (b) and signal (d) calculated
with particle level simulation (as explained in the text) for the case of
$L=10^{34}cm^{-2}s^{-1}$.}
\label{isol_hl}
\end{figure}

Assuming a charged track reconstruction efficiency 0.9 for tracks 
with $p_{t}>$ 0.9 GeV, we reproduced at the particle-level simulations 
the efficiency of the tracker isolation criterion obtained with full 
simulation and pattern recognition: $(2.7 \pm 0.2) \times 10^{-2}$ at the 
particle level as compared to $(3.0 \pm 0.2) \times 10^{-2}$  with full 
simulation and pattern recognition. This then allows us to obtain the 
efficiency of the tracker isolation criterion for the high luminosity case 
where we didn't make the full detector simulation.
Figures~\ref{isol_ll}a,c
and Figure~\ref{isol_hl}a,c show the tracker isolation parameter - the
number of charged tracks with  $p_{t} >$ 0.9 GeV in a cone
$\Delta R~=~0.5 \times \Delta R_{\mu \mu}~+~0.4$ around the di-muon 
momentum direction obtained with particle-level simulation as explained
above for the case of low (Figure~\ref{isol_ll}a,c) and high 
(Figure~\ref{isol_hl}a,c) luminosity.

\textbf{iii) secondary vertex selection}

\begin{figure}[htbp]
\centering
\resizebox{160mm}{160mm}
{\includegraphics{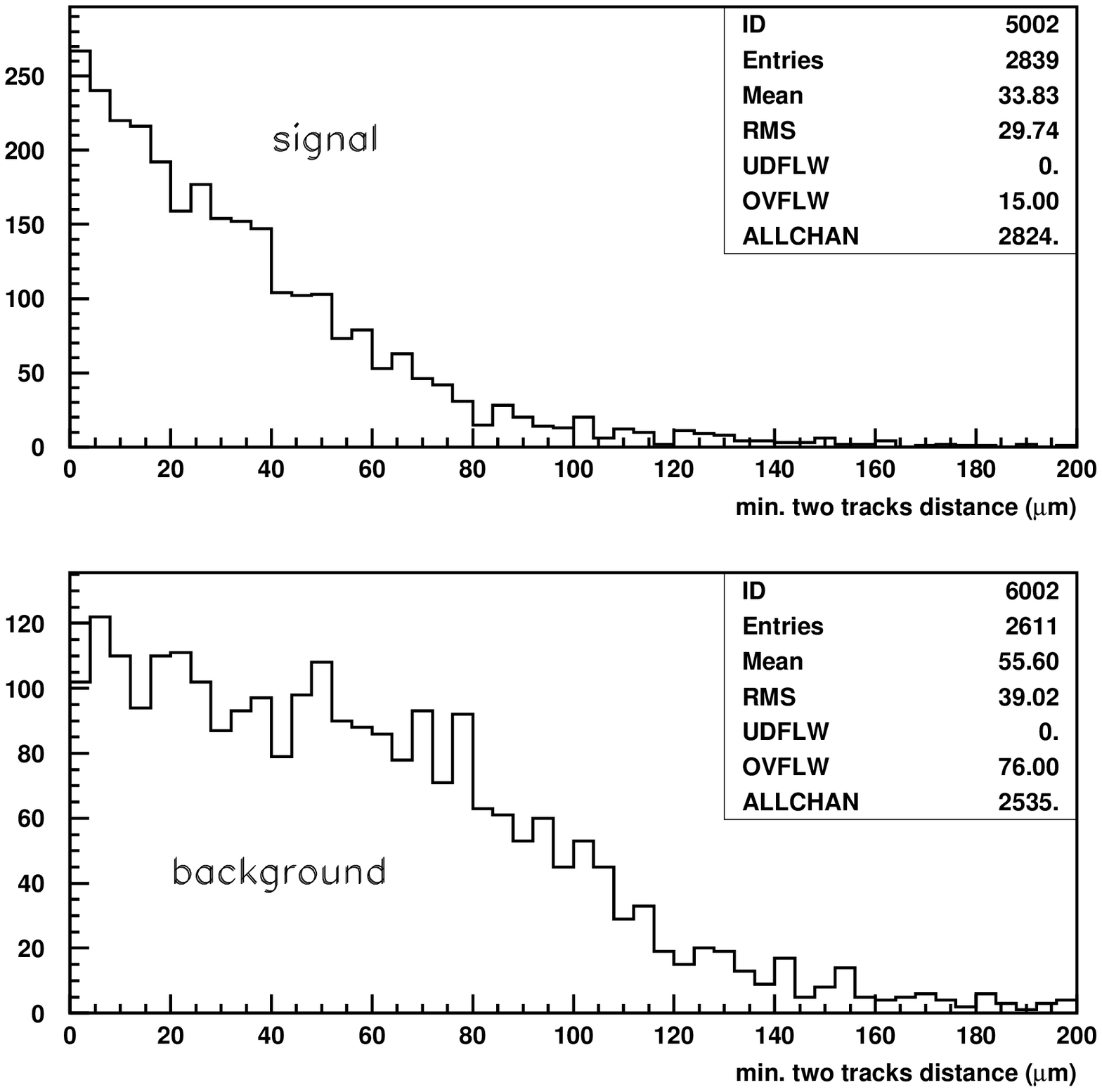}}
\caption
{Minimal distance in space between two muon tracks for the signal and 
background}
\label{m2d}
\end{figure}

\begin{figure}[htbp]
\centering
\resizebox{160mm}{160mm}
{\includegraphics{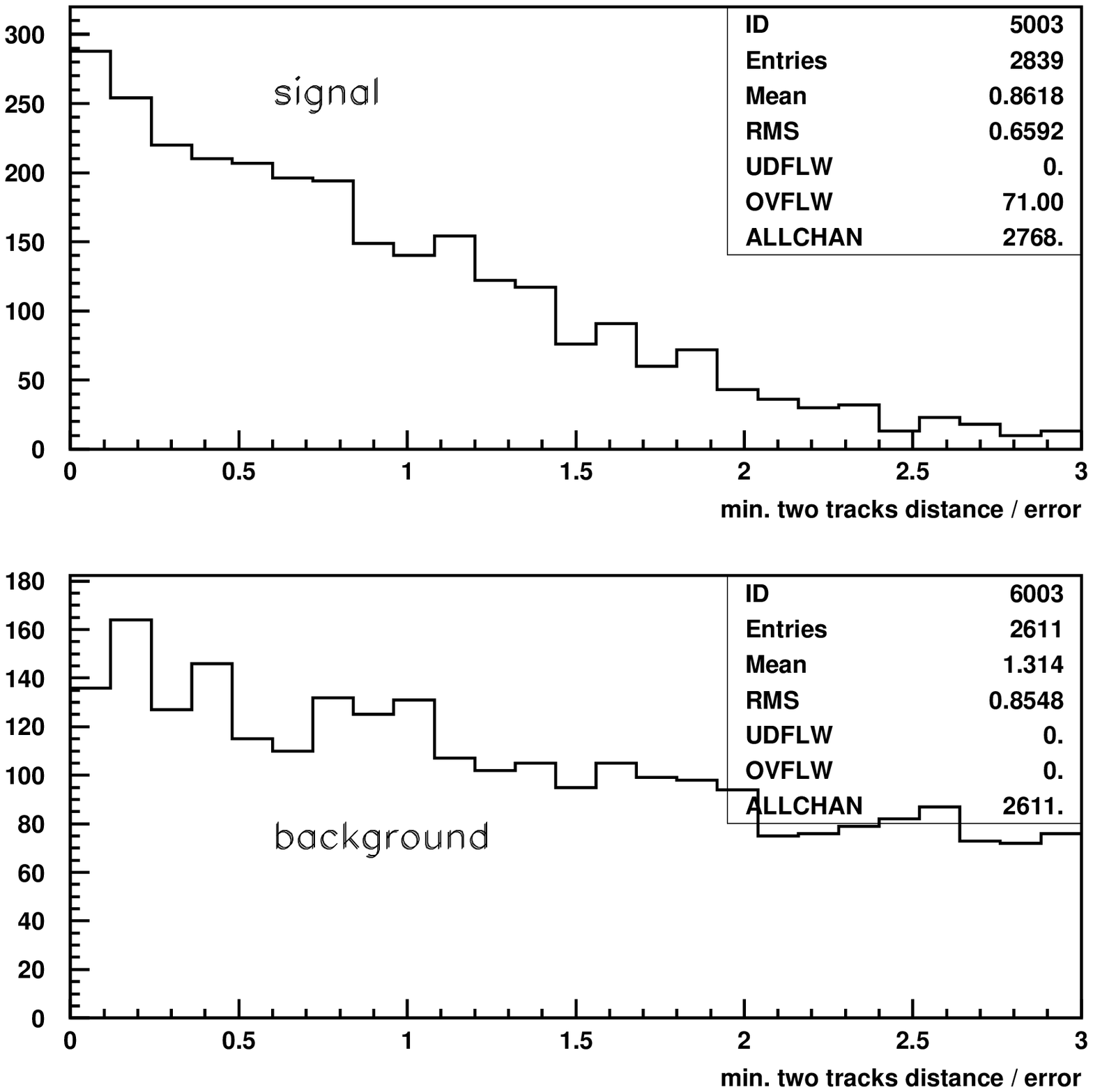}}
\caption
{Minimal distance in space between two muon tracks divided by the error of 
this value for the signal and background}
\label{m2d_rel}
\end{figure}

\begin{figure}[htbp]
\centering
\resizebox{160mm}{160mm}
{\includegraphics{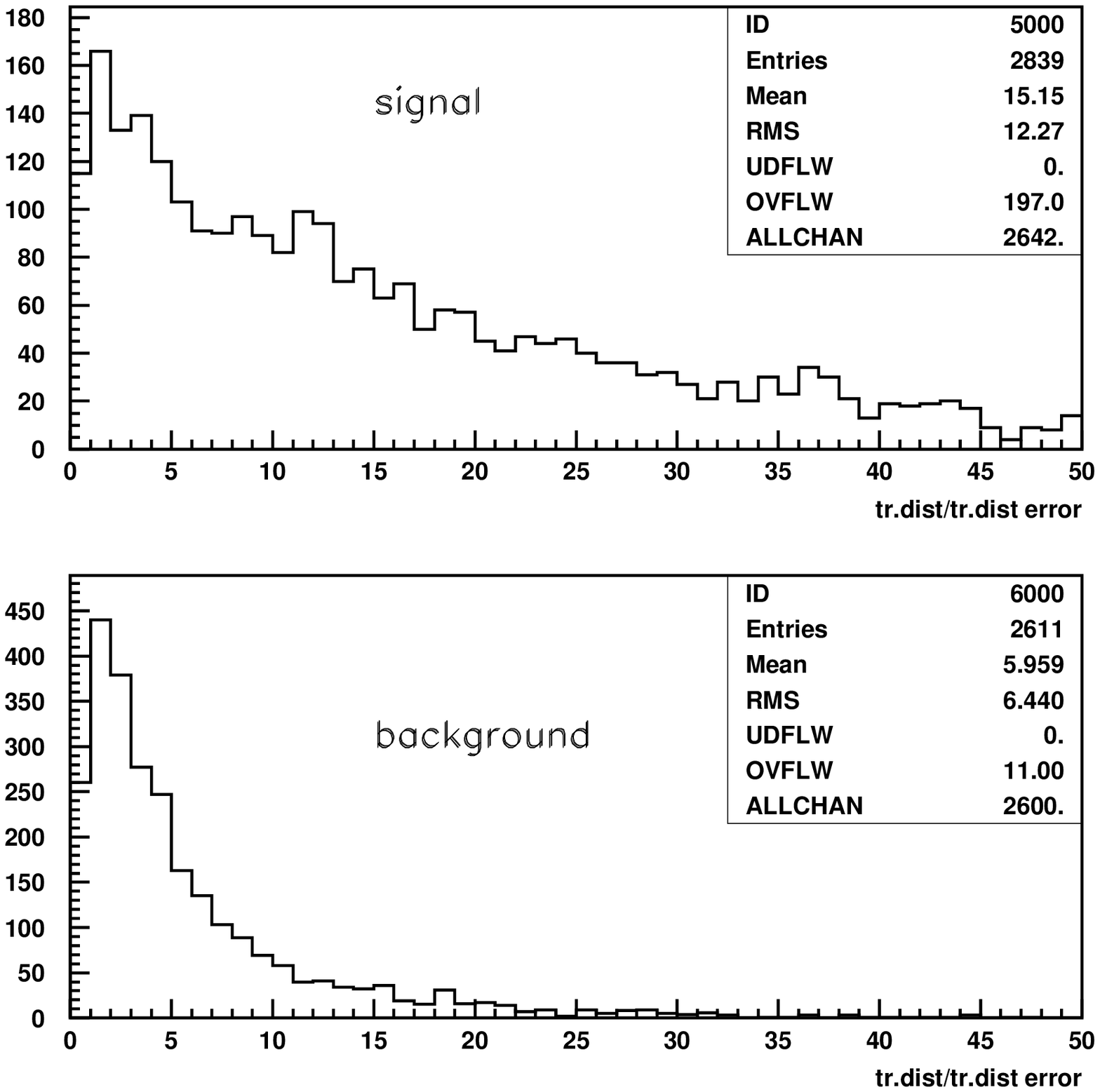}}
\caption
{Transverse distance (flight path) between secondary and primary vertex 
divided by the error of this quantity for signal and background}
\label{vrt_rel}
\end{figure}

\begin{figure}[htbp]
\centering
\resizebox{160mm}{160mm}
{\includegraphics{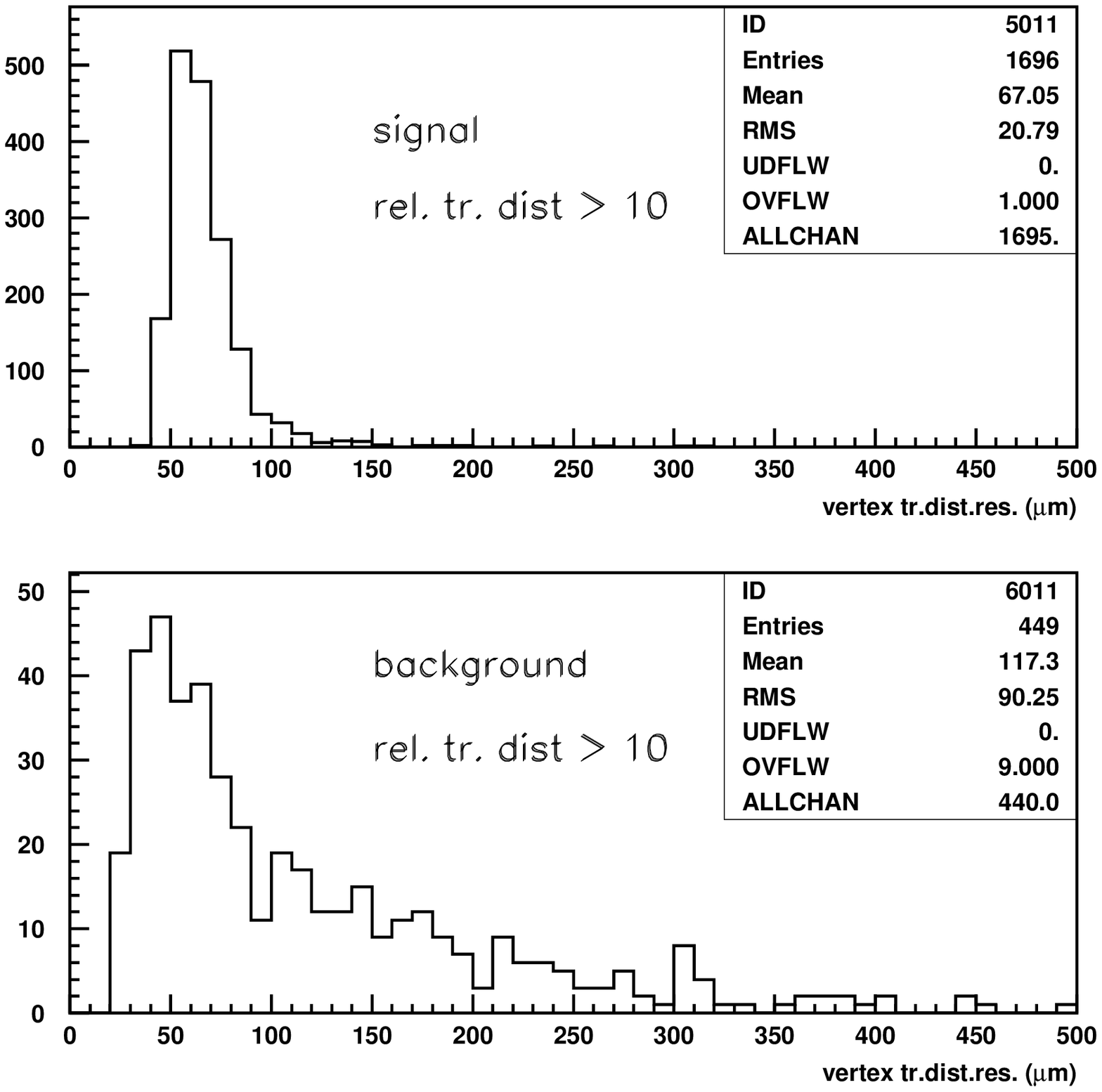}}
\caption
{Error on the transverse distance (flight path) between primary and 
secondary vertices for the signal and background}
\label{vrt_err}
\end{figure}

\begin{figure}[htbp]
\centering
\resizebox{160mm}{160mm}
{\includegraphics{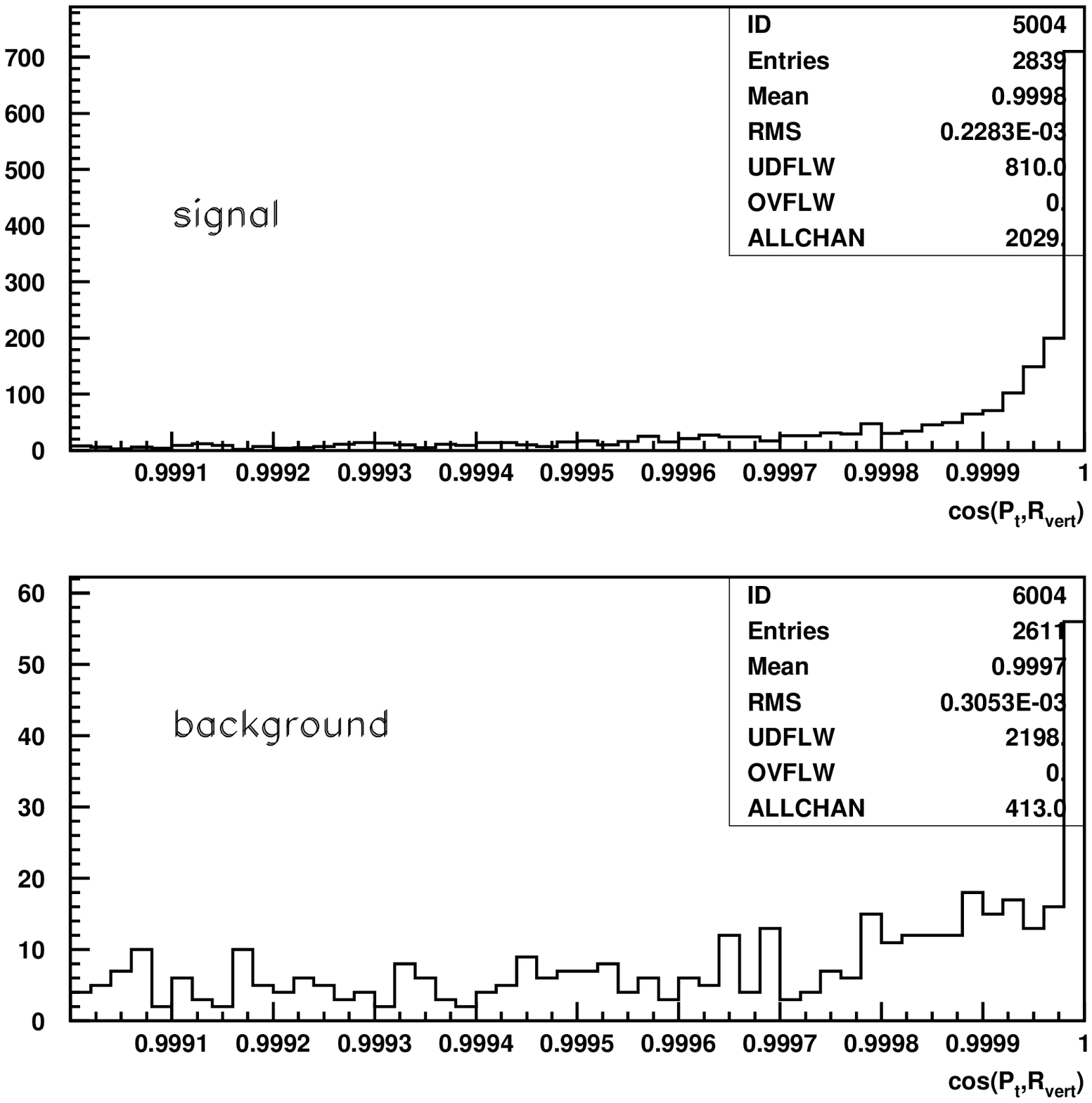}}
\caption
{Cosine of the angle in the transverse plane between the vector pointing
from the primary to secondary vertex (flight path) and the dimuon transverse 
momentum}
\label{cos_pr}
\end{figure}

A third set of cuts is based on the secondary vertex reconstruction
quality and primary-secondary vertex separation.
The minimal cuts on the reconstructed secondary vertex applied at the event 
reconstruction level allows to keep almost all signal events ($\sim
95$\%), but the background rejection factor is then about 4. 
The di-muon vertex reconstruction algorithm
provides however a number of parameters which can be used to improve the 
vertexing rejection power.  After the preliminary analysis, 5 variables were 
chosen:

\begin{enumerate}

\item M2D - minimal distance of approach in space between the two tracks to 
create the secondary vertex seed (Figure~\ref{m2d}).

\item M2D\_rel - relative minimal distance between the two tracks 
to create the secondary vertex seed (Figure~\ref{m2d_rel}).

\item VTR\_rel - relative transverse distance (transverse flight path) of 
the reconstructed secondary vertex to the primary vertex (by relative we 
mean here the variable measured in terms of
its errors ). It is shown in Figure~\ref{vrt_rel}

\item VTR\_err - absolute error on the reconstructed secondary vertex 
in the transverse plane (Figure~\ref{vrt_err}). 

\item COS\_pr - cosine of the angle between the vector pointing to the
secondary vertex position (flight path direction) and the two-tracks 
momentum ($B_{s}^{0}$ momentum) in the transverse plane (Figure~\ref{cos_pr}).
\end{enumerate}

To reject effectively the false background vertices one has to 
select high quality reconstructed secondary vertices which are rather far 
away from the primary interaction vertex in the transverse plane.
Optimising signal observability, we find that a rejection 
factor of order $10^{-4}$ can be obtained. The following set of cuts
allows to achieve such a rejection:    

\begin{enumerate}
\item $VTR\_rel \geq 12$ (as we checked this cut is nearly equivalent to a
sharp cut on the vertex distance in the transverse plane of about 820 $\mu$, 
however, we prefer to use a cut on a variable calculated on an event by event 
basis) 
\item $COS\_pr \geq 0.9997$ 
\item $VTR\_err \leq 80 \mu$ 
\item $M2D \leq 50 \mu$ 
\item $M2D\_rel \leq 2$ 
\end{enumerate}

This in fact allows to eliminate ALL events from the 10K background sample 
for the low luminosity tracker geometry, still keeping about 30 \% of the 
signal. We conclude that the vertexing rejection power is better than
$2.3 \times 10^{-4}$ at the 90\% C.L. level.

To obtain the same rejection power for the high luminosity tracker geometry
one has to use tighter cuts, in particular changing the cut on the relative 
transverse distance  $VTR\_rel \geq 12$ to $VTR\_rel \geq 15$ 
we can still reject all background events. However, a factor of 
$\approx 1.8$ in signal efficiency is lost. Other combinations of cuts 
investigated for the high luminosity geometry lead to comparable loss 
in signal efficiency.      

\subsection{Selection criterion based on calorimeters}

We have also included calorimeter isolation criteria to get an additional
background suppression factor. We required no transverse energy in the ECAL 
plus HCAL above 4 GeV (6 GeV) at $L=10^{33}cm^{-2}s^{-1}$ 
($L=10^{34}cm^{-2}s^{-1}$) in the same cone as for the tracker isolation.
An additional rejection factor of $2.3 \pm 0.2$ (both for low and high
luminosity) to the tracker isolation only has been obtained at the 
particle-level simulation and checked on the restricted sample of background 
events with full GEANT simulation. The efficiency of this additional 
criterion for the signal is $0.94 \pm 0.08$ for low luminosity and
$0.91 \pm 0.08$ for the high luminosity case.  The isolation energy 
distribution in the calorimeter obtained at particle level is shown in 
Figure~\ref{isol_ll}b,d for low luminosity running and in 
Figure~\ref{isol_hl}b,d for the high luminosity case.

\subsection{Correlation of the mass and isolation selections}

We have checked the correlation between the isolation of the dimuon pair 
(in the tracker and calorimeter) and the dimuon mass, to check whether it 
was correct to evaluate independently the efficiency of the mass cut and the 
isolation criteria.

\begin{figure}[htbp]
\centering
\resizebox{150mm}{230mm}
{\includegraphics{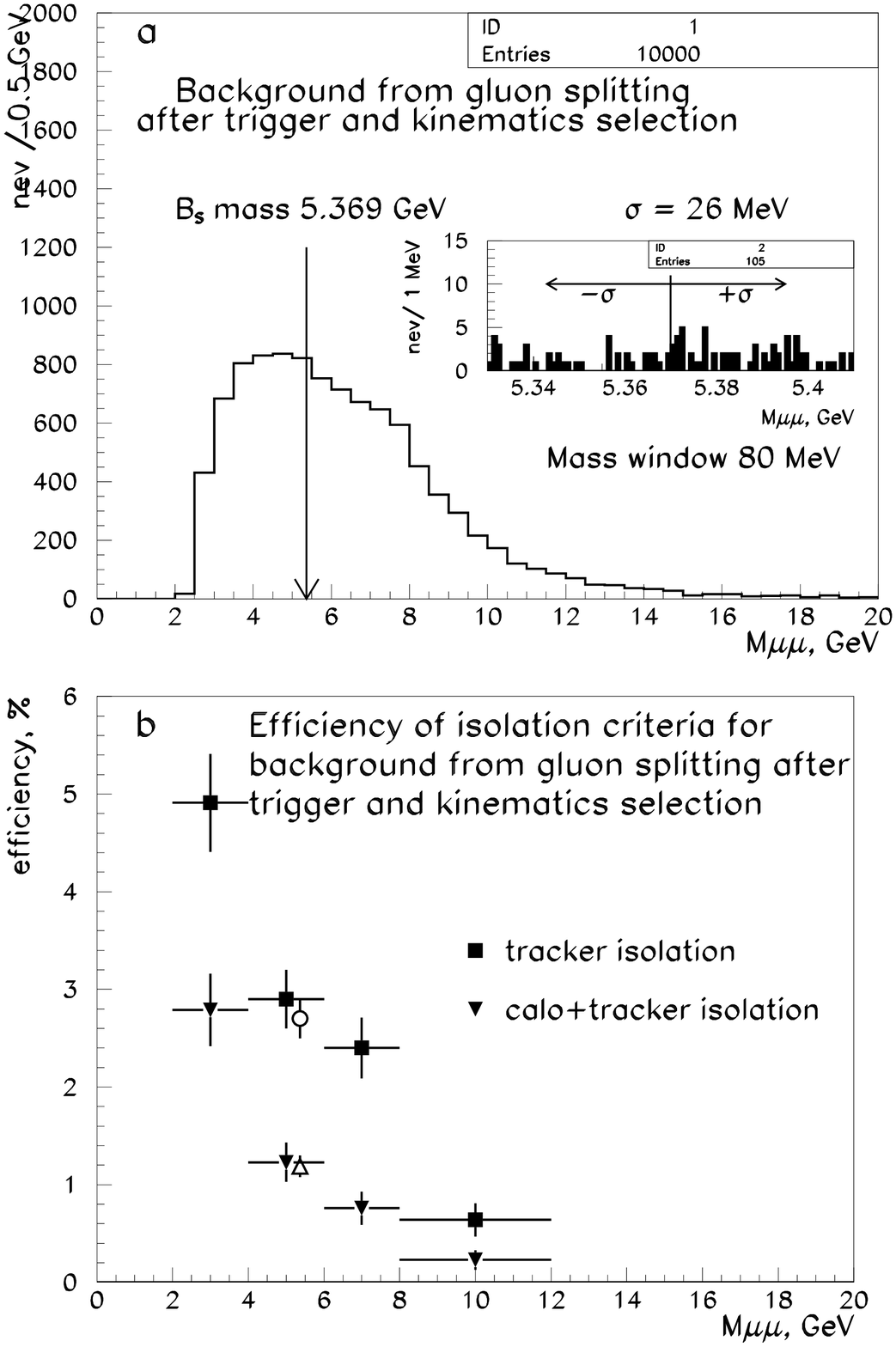}}
\caption
{a) - dimuon mass spectrum for the background from gluon splitting after
the trigger and kinematics selection. A mass window 80 MeV for the signal 
search is shown together with the signal resolution in the insertion. 
b) - efficiency
of the isolation criterion with the tracker only and with the tracker and 
calorimeter combined for the background from gluon splitting, after the 
trigger and kinematics selection.}
\label{isol_mass_corr}
\end{figure}

In Figure~\ref{isol_mass_corr}b the efficiency of the isolation criteria 
applied to different regions of di-muon mass is shown. One can clearly 
see a correlation. However, the efficiency of the isolation criterion 
obtained with full 
statistics over the whole di-muon mass interval 
(shown in the Figure~\ref{isol_mass_corr} b as open marks) coincides within 
the statistical errors with the one obtained for the proper interval of 
mass 4-6 GeV where the $B_{S}$ mass is located. One can also see in 
Figure~\ref{isol_mass_corr} b that the calorimeter isolation gives a 
sufficient additional suppression factor over the entire interval of dimuon 
mass.

\section{Conclusion}

Table 1 summarised the efficiency of the selection criteria and the number
of signal $B^{0}_{S} \rightarrow \mu \mu$
and background events after successive selection steps, for one
year running at both low and high luminosity. Due to insufficient Monte-Carlo
statistics we give for the number of background events an upper 
limit at 90\% C.L. As one can obtain from Table 1, the 
$B^{0}_{S} \rightarrow \mu \mu$ decay should be seen with significance 
4 after 3 year of running at a luminosity of
$L=10^{33}cm^{-2}s^{-1}$ even if the background were
underestimated by a factor 2. With $3 \times 10^{4} pb^{-1}$ data at low 
luminosity ($10^{33}cm^{-2}s^{-1}$) and $10^{5}pb^{-1}$ at high luminosity
($10^{34}cm^{-2}s^{-1}$) significance 6.3 can be achieved. 
This more optimistic expectation when 
compared to our previous study \cite{bs_old} comes from the more 
sophisticated algorithms for the track and vertex reconstruction and 
the tighter vertex selection criteria. 

Taking into account the $B^{0}/B_{S}^{0}$ = 0.40/0.11 production ratio and
the expected Standard Model 
$Br(B^{0}_{d} \rightarrow \mu \mu)=(1.5 \pm 0.9) \times 10^{-10}$, 
we also estimated that we should get for one year running at low luminosity 
$(1.1 \pm 0.7)$ of $B^{0}_{d} \rightarrow \mu \mu$ decays again on essentially
no background.

We should mention again that in the evaluation of the number of the signal and 
background events we have assumed 100 \% efficiency of the High Level 
Triggers for this di-muon channel. This might not be exactly the case 
although only a minor loss is expected, but a special study of the High Level 
Triggers efficiency for this mode is now under way.
Also we assumed full efficiency for tracker-muon chamber muon track matching; 
preliminary studies \cite{genchev} indicate that it is $\geq$90 \%.

\begin{table}[htb]
\begin{center}
{Table 1. Selection efficiencies and number of signal and background
events before and after selections for $10^{4}pb^{-1}$ running at 
low ($L=10^{33}cm^{-2}s^{-1}$) and $10^{5}pb^{-1}$ running at
high ($L=10^{34}cm^{-2}s^{-1}$) luminosity.}

\medskip

\begin{tabular}{|c|c|c|} 
\hline
  & Signal & Background \\
\hline
\hline 
number of events after trigger and kinematics selections & 
$66$ &  $2.9 \times 10^{7}$ \\
\hline
\hline
tracker isolation. Low luminosity &
$0.49$ & $3.0 \times 10^{-2}$ \\
\hline
tracker isolation. High luminosity &
$0.34$ & $2.0 \times 10^{-2}$ \\
\hline
tracker+calo isolation. Low luminosity &
$0.46$ & $1.3 \times 10^{-2}$ \\
\hline
tracker+calo isolation. High luminosity &
$0.31$ & $0.87 \times 10^{-2}$ \\
\hline
$2-\mu$ rec. + sec.vertex selections. Low luminosity &
$0.32$ & $\leq 2.3 \times 10^{-4}$ \\
\hline
$2-\mu$ rec. + sec.vertex selections. High luminosity &
$0.18$ & $\leq 2.3 \times 10^{-4}$ \\
\hline
mass window 80 MeV &
$0.72$ & $1.1 \times 10^{-2}$ \\
\hline
\hline
number of events after cuts. Low luminosity &
$7.0$ & $\leq 1.0$ at 90\% C.L. \\
\hline
number of events after cuts. High luminosity &
$26.0$ & $\leq 6.4$ at 90\% C.L. \\
\hline

\end{tabular}
\end{center}
\end{table}
 
\section{Acknowledgements}

We are thankful to D. Denegri, Y. Lemoigne and all members of the CMS B-physics
group for discussions and comments. We specially thank to D. Denegri
for careful reading of this note and very useful corrections.

\newpage
\begin{figure}[htbp]
\centering
\resizebox{150mm}{180mm}
{\includegraphics{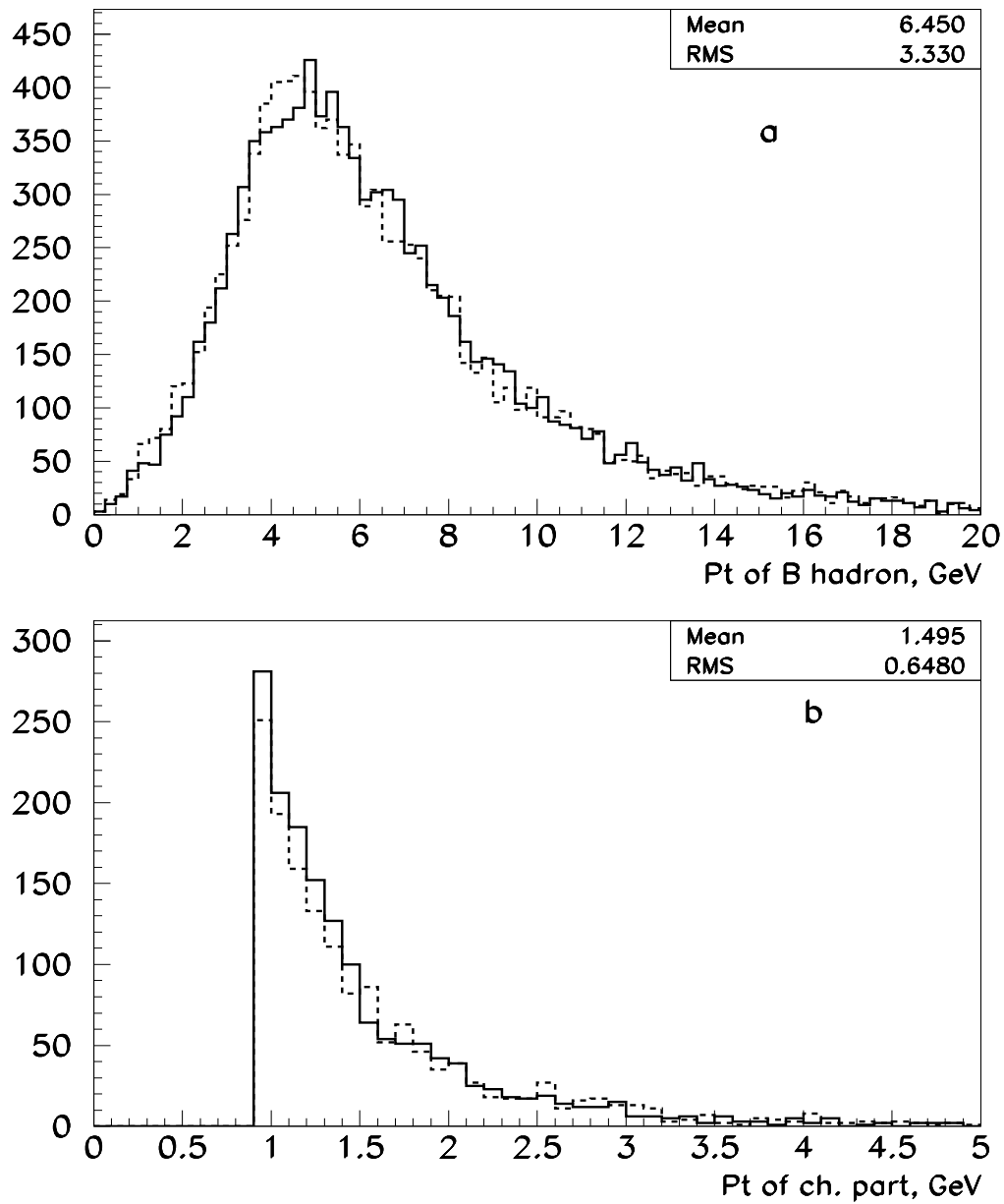}}
\caption
{$p_{t}$ spectra of B hadron (a) and charged particles (b) (not including
decay products of B-hadron) in cone 0.4 around B-hadron for different choice
 of fragmentation function in PYTHIA5.7. Solid line - PYTHIA5.7 default 
MSTJ(11)=4, dashed line - Peterson fragmentation MSTJ(11)=3}
\label{fragm_func}
\end{figure}
\end{document}